# Partial Identification of Answer Reviewing Effects in Multiple-Choice Exams

Yongnam Kim

University of Missouri–Columbia

Department of Educational, School & Counseling Psychology

16 Hill Hall, Columbia, MO 65211

ykcpb@missouri.edu

October 14, 2019

**ACKNOWLEDGEMENTS**

The previous version of this manuscript has been submitted to (12/21/2018) and rejected by (12/31/2018) *Educational Measurement: Issues and Practice*, submitted to (01/01/2019) and rejected by (01/02/2019) *Educational and Psychological Measurement*, and submitted to (01/02/2019) and rejected by (01/19/2019) *Applied Psychological Measurement*. The final version of the manuscript has been published in *Journal of Educational Measurement* (https://onlinelibrary.wiley.com/doi/full/10.1111/jedm.12259). The author thanks Felix Elwert and James Wollack for helpful comments on the previous version of the manuscript. James Wollack also generously provided his data set for use in this manuscript.

**ABSTRACT**

Does reviewing previous answers during multiple-choice exams help examinees increase their final score? This article formalizes the question using a rigorous causal framework, the potential outcomes framework. Viewing examinees' reviewing status as a treatment and their final score as an outcome, the article first explains the challenges of identifying the causal effect of answer reviewing in regular exam-taking settings. In addition to the incapability of randomizing the treatment selection (reviewing status) and the lack of other information to make this selection process ignorable, the treatment variable itself is not fully known to researchers. Looking at examinees' answer sheet data, it is unclear whether an examinee who did not change his or her answer on a specific item reviewed it but retained the initial answer (treatment condition) or chose not to review it (control condition). Despite such challenges, however, the article develops partial identification strategies and shows that the sign of the answer reviewing effect can be reasonably inferred. By analyzing a statewide math assessment data set, the article finds that reviewing initial answers is generally beneficial for examinees.

KEYWORDS: Answer reviewing, answer changing, multiple-choice exam, potential outcomes, causal inference

INTRODUCTION

Since the 1920s, measurement researchers have investigated how examinees review and change their answers while taking multiple-choice exams (e.g., Archer & Pippert, 1962; Benjamin, Cavell, & Shallenberger, 1984; Edwards & Marshall, 1977; Jeon, Boeck, & van der Linden, 2017; Lehman, 1928; Liu, Bridgeman, Gu, Xu, & Kong, 2015; Lynch & Smith, 1972; Matthews, 1929; McMorris, DeMers, & Schwarz, 1987; Pagni et al., 2017; Reile & Briggs, 1952; Skinner, 1983; van der Linden, Jeon, & Ferrara, 2011; Wainscott, 2016). Despite this long research history, however, some questions are still under debate. One such question is, *"Is changing initial answers generally beneficial or harmful?"* While many researchers who apply the traditional research method for this question (comparing the proportions of examinees' answer changing patterns) have argued that changing answers is beneficial (e.g., Benjamin et al., 1984; Bridgeman, 2012; Liu et al., 2015; Lynch & Smith, 1972), van der Linden et al. (2011) recently have claimed that the traditional method fails to control for examinees' ability levels, which will confound the relation between answer changing and final score, and their advanced item response theory (IRT) models approach "reveal[s] substantial losses due to changing initial responses for all ability levels" (van der Linden et al., 2011, p. 380). However, they later acknowledged that their empirical analysis was flawed and thus the conclusion remains inconclusive (see Bridgeman, 2012; Erratum of van der Linden et al., 2011).

A related but substantively different question under debate is, *"Should examinees review their previous answers?"* Whereas examinees cannot be forced to change their answers (e.g., suppose that an examinee was told, "you must change your first answer"), it is reasonable to give them an *opportunity* to change their initial answers such that the final answer changing decision is up to the examinees. Thus, unlike answer changing, answer reviewing is subject to policy evaluation (i.e., whether examinees should be allowed to review the previously answered items) and many educators and test administrators have wondered about the consequence of the reviewing policy.[1] This evaluation is crucial because it will directly affect how they design the test protocols and procedures. In particular, regarding computerized adaptive tests (CATs), item reviewing, which might distort the adaptive algorithm, has been an important issue in the recent measurement literature (e.g., Han, 2013; Jeon et al., 2017; Liu et al., 2015; Luecht & Nungester, 1998; Wainer, 1993; Wise, Finney, Enders, Freeman, & Severance, 1999; van der Linden et al., 2011).

Kim (2019) recently provided a new perspective that reframes the debate using the *potential outcomes* approach developed in the causal inference literature (Holland, 1986; Rubin, 1974, 1978). He separately defined the different subgroups' answer changing effects and showed that changing initial answers is beneficial to examinees who changed their answers but harmful to those who retained their answers.[2] He also pointed out that answer reviewing and answer changing are two separate processes (because an examinee may not change the initial answer even though he or she reviewed it) and whether examinees would have gains by *reviewing* their answers (i.e., answer reviewing effect) cannot be predicted from whether examinees would have gains by *changing* their answers (i.e., answer changing effect). However, Kim (2019) focused on the answer changing effect in that study instead of the answer reviewing effect, and how to investigate the answer reviewing effect remained unclear. Although he suggested a randomized experiment where researchers can manipulate examinees' reviewing status, as in Vispoel's (2000) study, such a randomizing process is hard to implement in a real-world setting.

---

[1] According to Pearl's (2014) distinction, answer changing effects can be viewed as being subject to scientific knowledge (because there is no proper way to implement answer changing) while answer reviewing effects can be viewed as being subject to policy evaluation.

[2] The former is known as the average treatment effect on the treated (ATT) while the latter is known as the average treatment effect on the untreated (ATU). See Kim (2018) for how they can be identified despite the fact that some examinees retained their answers and thus their counterfactual answer correctness if they had changed the answers remains unknown. Similar definitions and logics are used in this article. See the next section.

The purpose of this article is to apply the potential outcomes approach to infer the answer reviewing effect in a regular exam-taking setting. In this setting, examinees self-select whether they would review their answers for a specific item or not, and researchers cannot intervene in this process. Researchers can access examinees' answer sheets once the exam is completed. Is it possible to infer the causal effect of answer reviewing given only the answer sheet data? From a conventional causal inference point of view, it seems almost impossible, because the examinees' reviewing status, which is the treatment variable, is not fully known to researchers. From the answer sheet data alone, it is unclear whether an examinee who did not change his or her answer for a specific item reviewed it but retained the initial answer (treatment condition) or chose not to review it (control condition). Nonetheless, while it may not be possible to estimate the magnitude of the effect, this article shows that it is possible to determine the sign of the answer reviewing effect from the observed answer sheet data, thereby enabling us to learn whether answer reviewing is generally beneficial or harmful. The rationale and the conditions under which the sign of the answer reviewing effect can be inferred will be explicitly formalized with potential outcomes.

The rest of this article is organized as follows. First, the basic setup, notation, and definitions are introduced. Using potential outcomes, different types of causal effects of answer reviewing are defined. Next, the main section shows how to impute the missing potential outcomes and derives analytic formulas of the defined answer reviewing effects. Using the derived partial identification strategies (e.g., assumption-free bounds), the subsequent section analyzes a statewide math assessment data set and shows that reviewing answers is generally beneficial to examinees. The article concludes with a discussion of the theoretical and practical implications of the findings.

## PRELIMINARY CONCEPTS

**Setup**

This article investigates the answer reviewing effect in a regular exam-taking setting. $N$ examinees take a multiple-choice exam, consisting of $J$ items. Each item has more than two (typically four or five) alternatives, and only one of them is correct. The focus is on paper-and-pencil tests where examinees' answers are marked on an optical answer sheet such as a bubble sheet, but the principle may also be applied to computer-based tests. Once their first answers are marked on the answer sheets, examinees self-select whether they will review the answers or not. Note that this means that any mental processes before marking the initial answers are not considered answer reviewing in this article.[3] Depending on the remaining time and other factors like their confidence, fatigue, or anxiety at the time, it is possible that they may choose not to review a marked answer. Although some studies assume that all examinees review their initial answers during exams (e.g., Jeon et al., 2017; van der Linden et al., 2011; van der Linden & Jeon, 2012), it is rather implausible and this article considers that examinees self-select whether they review their initial answers or not.[4] Once examinees review their answers, they may switch them

---

[3] Consequently, under paper-and-pencil test conditions, two different types of reviewing decisions cannot be distinguished. Suppose a student first marked 'A' and then immediately realized he or she mismarked it, and so changed it to 'B' before moving on to the next item. This examinee's decision to review the item may be qualitatively different from other reviewing decisions that are made by going back to the item after solving other items. Such a distinction may be possible in computer-based tests where examinees' every activity is recorded as a log file, but not in paper-and-pencil tests.

[4] If all examinees review their initial answers, the investigation of the causal effect of answer reviewing is straightforward because then examinees' initial answers correspond to their

by erasing the previous choice and choosing an alternative. It is important to note that examinees might not change their marked answers even though they reviewed the answers. It is accepted that examinees who change their answers do so only after careful consideration—that is, there is no unconscious answer changing without reviewing. Once the exam is completed, all examinees' answer sheets are scanned by an optical scanner that can detect both previously erased and newly selected answers. Researchers who investigate the answer reviewing effect can access the scanned answer sheet data in which all examinees' first and final answers are recorded.[5] The answer keys of all *J* items are given, but no other information about examinees (e.g., sex, IQ score, etc.) is available to researchers.

**Notation and Definitions**

This article uses *i* as examinee index and *j* as item index. Let $F_i^j$ denote whether examinee *i*'s first answer on item *j* is correct ($F_i^j = 1$) or incorrect ($F_i^j = 0$). $T_i^j$ denotes the answer reviewing status such that examinee *i* reviews his or her first marked answer ($T_i^j = 1$) or

---

potential control outcomes and their final answers correspond to their potential treatment outcomes. See the following subsection for explanations about potential outcomes.

[5] It is possible that an examinee may switch his or her answers multiple times. Especially when using bubble sheets, this multiple answer changing may cause a complexity in investigating the answer reviewing effect. For example, an examinee first chose 'A,' but erased it and switched to 'B' (1st change). Later, the examinee may switch back to 'A' (2nd change), which is the initial answer. If this is the case, by scanning the bubble sheet, researchers may mistakenly believe that 'B' was the initial answer. This issue has not been explicitly discussed in the literature and researchers typically assume that examinees' first and final answers can be correctly detected (or believe that answer changing occurs at most one time in each item). This article makes the same assumption as with many other studies (e.g., Benjamin et al., 1984; Jeon et al., 2017; van der Linden et al., 2011) and does not take this complexity into account. Although this omission may be problematic, the findings of this article can be still useful because, first, the ratio of such multiple corrections is frequently very low (e.g., 2.5% of the total number of answer changes in Mathews, 1929), and second, this error does not occur if computer-based tests are used because examinees' every response is automatically recorded in a log file.

not ($T_i^j = 0$) on item $j$. $Y_i^j$ denotes whether $i$'s final answer on the item is correct ($Y_i^j = 1$) or incorrect ($Y_i^j = 0$). From a causal inference point of view, $T$ can be viewed as a treatment, $Y$ an outcome, and $F$ a covariate. In addition, using the potential outcomes framework, this article defines two potential outcomes corresponding to the binary treatment status. The *potential treatment outcome* $Y_i^j(1)$ is defined as the *hypothetical* final answer correctness if $i$ had reviewed the first answer on item $j$, where the parenthetical value indicates the treatment condition, $T_i^j = 1$. The *potential control outcome* $Y_i^j(0)$ is defined as the hypothetical final answer correctness if $i$ would *not* have reviewed the first answer on item $j$, where the parenthetical value indicates the control condition, $T_i^j = 0$. Note that the potential outcomes $Y_i^j(1)$ and $Y_i^j(0)$ are conceptually different from the actual observed outcome $Y_i^j$.

The causal effect of answer reviewing for examinee $i$ on item $j$ is defined as the difference between the two potential outcomes:

$$\tau_i^j = Y_i^j(1) - Y_i^j(0). \tag{1}$$

This causal effect is referred to as the individual or unit-level causal effect. Average causal effects across examinees can be defined in several ways. First, the average treatment effect (ATE) of answer reviewing on item $j$ is defined as

$$ATE^j = E[\tau_i^j] = E[Y_i^j(1)] - E[Y_i^j(0)], \tag{2}$$

where the expectation is taken across all examinees. One may alternatively define the average causal effect for a subgroup of examinees. For example, the causal effect can be defined for a group of examinees who reviewed their answers (i.e., treated group). This type of causal effect is referred to as the average treatment effect on the treated (ATT) and is defined as

$$ATT^j = E[\tau_i^j \mid T_i^j = 1] = E[Y_i^j(1) \mid T_i^j = 1] - E[Y_i^j(0) \mid T_i^j = 1], \tag{3}$$

and the expectation is taken across examinees who reviewed their answers. One may be interested in the other subgroup of those who did *not* review their answers (i.e., control group). The average treatment effect on the untreated (ATU) is defined as

$$ATU^j = E[\tau_i^j | T_i^j = 0] = E[Y_i^j(1) | T_i^j = 0] - E[Y_i^j(0) | T_i^j = 0], \qquad (4)$$

where the expectation is taken across examinees who did not review their answers. As such, the causal effect of answer reviewing can be differently defined depending on researchers' interests.

**Identification Problems**

The purpose of the current analysis is to identify the defined causal effects of answer reviewing. In the causal inference literature, it is well known that the unit-level causal effect, defined in Equation (1), is generally not identified because only one of the two potential outcomes is realized in reality, while the other is missing (i.e., the fundamental problem of causal inference; Holland, 1986). For example, if examinee $i$ decided not to review the first answer on item $j$, then his or her potential control outcome $Y_i^j(0)$ is realized, but the potential treatment outcome $Y_i^j(1)$ is not realized and remains unknown. Thus, the unit-level causal effect, $\tau_i^j = Y_i^j(1) - Y_i^j(0)$, cannot be computed. However, relying on the strong ignorability assumption (Rosenbaum & Rubin, 1983), the average causal effects, as defined in Equations (2) to (4), can be identified. The identification of the ATE, defined in Equation (2), requires that the treatment assignment (i.e., whether examinees review or not) is randomized, or more generally, is

*ignorable* conditional on a set of covariates. Formally, when $Y^j(1), Y^j(0) \perp T^j \mid X$, where $X$ denotes covariates, the average causal effects for item *j* can be identified.[6]

However, in real exam-taking settings, examinees self-select to review their answers and the factors affecting this decision often remain unmeasured. Not only may examinees' latent ability affect whether they decide to review the previously answered items (van der Linden at al., 2011; van der Linden & Jeon, 2012), but so too may other unmeasured non-cognitive factors such as personality, anxiety, or fatigue.[7] Although one might try to measure such latent factors to deconfound the relation between answer reviewing and final score, it is unlikely that all of them can be perfectly measured such that the self-selected reviewing decision becomes ignorable by conditioning on the measured covariates.[8] Indeed, this article already assumed that such information is unavailable. Thus, the identification of the average causal effects of answer reviewing in the setup of this article is more challenging than in others, such as Vispoel's (2000) study where the reviewing condition was randomized by researchers.

There is one further difficulty that makes the identification of average causal effects almost impossible. From the marked answer sheets, researchers do not fully know examinees' reviewing status (i.e., treatment). For example, if examinee $i$'s first answer on item $j$ was 'A' but the final answer is 'B,' then the examinee must have reviewed the item ($T_i^j = 1$). However, if the initial answer was 'A' and the final answer is also 'A,' then it is not clear whether the examinee reviewed the item or not. He or she might have decided not to review the item ($T_i^j = 0$) or reviewed it ($T_i^j = 1$) but decided not to change the first answer. The goal of this article is to make a reasonable inference about the answer reviewing effect even when examinees' reviewing status is only partially known to researchers.

---

[6] To be precise, the identification of the average causal effects defined in Equation (3) and (4) indeed requires a weaker assumption than the strong ignorability. Instead of the conditional independence with respect to the joint distribution of the two potential outcomes, $Y^j(0) \perp T^j \mid X$ is required for ATTs and $Y^j(1) \perp T^j \mid X$ for ATUs, respectively.

[7] Another interesting case of non-randomized reviewing status occurs when multiple items exist. The decision to review one item may subsequently affect the decision to review another item because of, for example, the shortage of time. The approach developed in this article does not assume randomized reviewing status; therefore, the potential interference between each item's reviewing decision does not restrict the application of the approach.

[8] Alternatively, one may try to infer examinees' latent ability levels by relying on modeling assumptions (e.g., van der linden et al., 2011; van der Linden & Jeon, 2012). The credibility of the estimated answer reviewing effects will then depend on the validity of the specified models. Whether the estimated ability level will capture other non-cognitive factors that can confound the relation between answer reviewing and final score, should also be considered.

## IDENTIFICATION OF AVERAGE ANSWER REVIEWING EFFECTS

**Inferring Potential Outcomes**

In order to investigate the answer reviewing effect, this article builds on Kim (2019), who studied the causal effect of answer changing, instead of answer reviewing. Hereafter, the answer reviewing effect for a single item is considered and, for ease of notation, the item index $j$ is suppressed until real data with multiple items are analyzed in the next section. The provided answer sheet data produce two observed variables $F$ and $Y$, the first and the final answer correctness, respectively. There are four possible combinations of the pair of ($F$, $Y$): $WW$ ("wrong"-"wrong"), $WR$ ("wrong"-"right"), $RW$ ("right"-"wrong"), and $RR$ ("right"-"right"). For $WR$ and $RW$ types, their reviewing status is directly inferred. Those who belong to either type have *changed* their answers; therefore, they must have reviewed the answers, $T_{i \in WR, RW} = 1$. However, the others who belong to either $WW$ or $RR$ may or may not have reviewed their answers and thus their reviewing status is not certain. Note that a group of examinees who belong to $WW$ but reviewed their previous answers may be identified from the answer sheet data if they switch their first wrong answers to other wrong answers. This specific group will be

TABLE 1.
*Observed first and final answer correctness, the inferred reviewing status, and the corresponding potential outcomes*

| Type | F | Y | T | Y(1) | Y(0) |
|------|---|---|---|------|------|
| WW   | 0 | 0 | 0 | (a)  | 0    |
|      | 0 | 0 | 1 | 0    | 0    |
| WR   | 0 | 1 | 1 | 1    | 0    |
| RW   | 1 | 0 | 1 | 0    | 1    |
| RR   | 1 | 1 | 0 | (b)  | 1    |
|      | 1 | 1 | 1 | 1    | 1    |

*Note.* $F$ = first answer correctness; $Y$ = final answer correctness; $T$ = answer reviewing status; $Y(1)$ = hypothetical answer correctness if an examinee had reviewed; $Y(0)$ = hypothetical answer correctness if an examinee would not have reviewed; $WW$ = "wrong"-"wrong" type; $WR$ = "wrong"-"right" type; $RW$ = "right"-"wrong" type; $RR$ = "right"-"right" type. (a) and (b) indicate two missing potential outcomes.

further discussed later (see $\kappa$ in Equation 9). Table 1 presents all possible combinations of $F$, $Y$, and $T$, where only WW and RR have two separate rows for distinguishing two different values of $T$, $T_{i \in WW, RR} = 0$ or $T_{i \in WW, RR} = 1$.

In order to identify the causal effects of answer reviewing, potential outcomes must be inferred. First, all examinees' potential control outcomes are directly inferred from observed data. If an examinee did not review his or her answers, the first answer for a given question would not have changed and thus would equal the final answer. Therefore, for any examinee $i$, the potential control outcome equals the first answer correctness, $Y_i(0) = F_i$. In contrast, inferring potential treatment outcomes is not that straightforward. Nonetheless, researchers can infer the potential treatment outcomes of those who reviewed their answers. If an examinee reviewed his or her first answer ($T_i = 1$), the potential treatment outcome $Y(1)$ is realized in reality and becomes the actual outcome $Y$. However, the potential treatment outcomes of those who did not review their answers ($T_i = 0$) are not realized and thus their potential treatment outcomes $Y(1)$ remain unknown and missing (instead, their potential control outcomes *Y*(0) are realized). In the causal inference literature, this principle is referred to as *consistency* (Robins, 1986), formally expressed as $Y_i = Y_i(1) \times T_i + Y_i(0) \times (1 - T_i)$.

The imputed potential outcomes, using both $Y_i(0) = F_i$ and the consistency, are presented in Table 1. Note that the potential treatment outcomes of those who did not review their answers remain unknown, denoted by (a) and (b). For those who belong to *WW* and did not review their answers ($T_i = 0$), their hypothetical final answer correctness if they had reviewed— that is, their potential treatment outcome *Y*(1)—may be right or wrong, (a) = 0 or 1. The same is true for those who belong to *RR* and did not review their answers, therefore, (b) = 0 or 1.

**Analytic Formulas for Answer Reviewing Effects**

In Table 2, the previous table is modified such that all possible values for (a) and (b) are displayed. All eight mutually exclusive groups are described in Table 2. For example, $WW_1$, which is a subgroup of *WW*, represents those who did not review their initial answers, $T_{i \in WW_1} = 0$, and even if they had reviewed them, would not have corrected them, $Y_{i \in WW_1}(1) = 0$. In contrast, $WW_2$, another subgroup of *WW*, represents those who did not review their initial answers, $T_{i \in WW_2} = 0$, but if they had reviewed them, would have corrected them, $Y_{i \in WW_2}(1) = 1$. Thus, unlike $WW_1$, $WW_2$ would have realized their mistakes and corrected them if they had reviewed. A similar distinction is applied to $RR_1$ and $RR_2$.

As all potential outcomes are imputed, the causal effect is directly computed by $\tau_i = Y_i(1) - Y_i(0)$. Table 2 shows that the causal effect of answer reviewing is heterogeneous

TABLE 2.
*Adding causal effects and group proportions to Table 1*

| Type | F | Y | T | Y(1) | Y(0) | $\tau_i$ | $P(\cdot)$ |
|---|---|---|---|---|---|---|---|
| WW | 0 | 0 | 0 | (a = 0) | 0 | 0 | $P(WW_1)$ |
|    | 0 | 0 | 0 | (a = 1) | 0 | +1 | $P(WW_2)$ |
|    | 0 | 0 | 1 | 0 | 0 | 0 | $P(WW_3)$ |
| WR | 0 | 1 | 1 | 1 | 0 | +1 | $P(WR)$ |
| RW | 1 | 0 | 1 | 0 | 1 | −1 | $P(RW)$ |
| RR | 1 | 1 | 0 | (b = 0) | 1 | −1 | $P(RR_1)$ |
|    | 1 | 1 | 0 | (b = 1) | 1 | 0 | $P(RR_2)$ |
|    | 1 | 1 | 1 | 1 | 1 | 0 | $P(RR_3)$ |

*Note*. $F$ = first answer correctness; $Y$ = final answer correctness; $T$ = answer reviewing status; $Y(1)$ = hypothetical answer correctness if an examinee had reviewed; $Y(0)$ = hypothetical answer correctness if an examinee had not reviewed; $\tau = Y(1) - Y(0)$; $P(\cdot)$ = group proportion; $WW$ = "wrong"-"wrong" type; $WR$ = "wrong"-"right" type; $RW$ = "right"-"wrong" type; $RR$ = "right"-"right" type.

such that it may be positive ($+1$), negative ($-1$), or zero. $P(\cdot)$ denotes the corresponding group proportion to each row in the entire population such that $P(WW) + P(WR) + P(RW) + P(RR) = 1$, where $P(WW) = P(WW_1) + P(WW_2) + P(WW_3)$ and $P(RR) = P(RR_1) + P(RR_2) + P(RR_3)$. From Table 2, the ATE of answer reviewing is expressed as the weighted average of the group causal effects:

$$ATE = P(WW_2) + P(WR) - P(RW) - P(RR_1). \tag{5}$$

And the ATT of answer reviewing is the causal effect only for those who reviewed their answers (i.e., $T_i = 1$) and is expressed as

$$ATT = \frac{P(WR) - P(RW)}{P(T=1)} = \frac{P(WR) - P(RW)}{P(WW_3) + P(WR) + P(RW) + P(RR_3)}, \tag{6}$$

where $P(T = 1)$ denotes the proportion of those who reviewed their answers. Similarly, the ATU of answer reviewing is expressed as

$$ATU = \frac{P(WW_2) - P(RR_1)}{P(T=0)} = \frac{P(WW_2) - P(RR_1)}{P(WW_1) + P(WW_2) + P(RR_1) + P(RR_2)}, \quad (7)$$

where $P(T=0)$ denotes the proportion of those who did not review their answers.

**Partial Identification of Answer Reviewing Effects**

Although the analytic formulas for the average answer reviewing effects are derived in Equations (5) to (7), such effects are not directly identified. The problem is that, although the four major groups $WW$, $WR$, $RW$, and $RR$ are observed and thus their group proportions are known, the subgroups $WW_1$, $WW_2$, $WW_3$, $RR_1$, $RR_2$, and $RR_3$ are latent and the corresponding group proportions cannot be inferred from the answer sheet data. For example, the ATT of answer reviewing cannot be computed by Equation (6) because researchers do not know $P(WW_3)$ and $P(RR_3)$. Nonetheless, one can identify the *sign* of the ATT because the numerators of the fractions in Equation (6) consist of the observed terms $P(WR)$ and $P(RW)$ and the denominators are positive $P(T=1) = P(WW_3) + P(WR) + P(RW) + P(RR_3) > 0$. Thus, if the proportion of those who changed their wrong answers to right answers is greater than the proportion of those who changed their right answers to wrong answers, $P(WR) > P(RW)$, then the average answer reviewing effect for those who reviewed their answers (ATT) is positive. Conversely, if the former is less than the latter, $P(WR) < P(RW)$, the ATT of answer reviewing is negative. Formally,

$$\text{sgn}(ATT) = \text{sgn}(P(WR) - P(RW)). \quad (8)$$

Similarly, the ATE of answer reviewing defined in Equation (5) is not identified because $P(WW_2)$ and $P(RR_1)$ are unknown. Still, it may be possible to reasonably infer its sign by deriving the bounded effect of the ATE. Note that the two latent group proportions, $P(WW_2)$ and

$P(RR_1)$, can be bounded. The former is bounded as $0 \leq P(WW_2) \leq P(WW) - \kappa$, where $\kappa$ denotes the observed proportion of examinees who switched their initial wrong answers to another wrong answers which can be identified from the answer sheet data. The proportion of the latter group is bounded as $0 \leq P(RR_1) \leq P(RR)$. Then, the bound on the ATE of answer reviewing is given by

$$P(WR) - P(RW) - P(RR) \leq ATE \leq P(WW) + P(WR) - P(RW) - \kappa, \qquad (9)$$

consisting of all observed proportions. Therefore, if the bound does not include zero, then the sign of the ATE is known. To be specific, if the lower bound of Equation (9) is greater than zero, then the sign of the ATE is positive. If the upper bound of Equation (9) is less than zero, then the sign of the ATE is negative. Although this assumption-free bound may allow researchers to infer the sign of the ATE of answer reviewing, as will be shown in the next section, this bound is rather wide and frequently contains zero so that it is generally hard to conclude whether the ATE of answer reviewing is positive or negative.

However, if one is willing to accept an assumption, the bound in Equation (9) can be further tightened and inferring the sign of the ATE may be easier. First, consider the examinees who belong to $WW_2$. As can be seen in Table 2, they chose a wrong answer at first, $F_{i \in WW_2} = 0$, and did not review it, $T_{i \in WW_2} = 0$. But, if they had reviewed it (counterfactual), they would have switched to the correct answer, $Y_{i \in WW_2}(1) = 1$. Next, consider those who belong to $RR_1$. They first chose a correct answer, $F_{i \in RR_1} = 1$, and did not review it, $T_{i \in RR_1} = 0$. But, if they had reviewed it (counterfactual), they would have switched the correct first answer to a wrong answer, $Y_{i \in RR_1}(1) = 0$. In many regular academic settings where examinees do their best to increase their scores, it is likely that the group $RR_1$ is relatively rarer than the group $WW_2$, $P(WW_2) > P(RR_1)$.

Although it is plausible in principle, this assumption is empirically untestable because it states a comparison between two unrealized counterfactuals. Researchers should assess its validity using subject-matter knowledge. If one is willing to accept this assumption, the bound in Equation (9) can be tightened as

$$P(WR) - P(RW) \leq ATE \leq P(WW) + P(WR) - P(RW) - \kappa, \qquad (10)$$

by correcting the lower bound using $P(WW_2) - P(RR_1) > 0$. Thus, under this assumption, if one finds $P(WR) > P(RR)$ from the answer sheet data, the ATE of answer reviewing becomes positive because the lower bound in Equation (10) is greater than zero.

Finally, the ATU is also not identified because all the group proportions in Equation (7) are unknown. In fact, the assumption-free bound on the ATU is given by

$$-1 \leq ATU \leq 1, \qquad (11)$$

which is not informative. Thus, the point estimate and even the sign of the causal effect of answer reviewing for examinees who did not review their answers (ATU) cannot be inferred from the answer sheet data alone. However, the assumption used for tightening the bound on the ATE can also allow researchers to infer the sign of the ATU. If the assumption $P(WW_2) > P(RR_1)$ is met, then the ATU is positive because $P(WW_2) - P(RR_1)$ is the numerator in Equation (7). Thus, the sign of the ATU of answer reviewing can be directly inferred under the assumption even without referring to empirical data.

Note that the proposed partial identification formulas for answer reviewing effects (i.e., Equations 8, 9, 10) do not rely on any covariates or the strong ignorability assumption (or the randomization of the answer reviewing status). Also, no parametric modeling assumptions (e.g., IRT models or linear systems having constant effects) have been made. Based on only clear definitions of the causal effects of answer reviewing, expressed with potential outcomes, and the consistency, together with $P(WW_2) > P(RR_1)$, it is possible to infer at least the sign of answer reviewing effects even in the presence of confounding, caused by the non-randomized answer reviewing choice.

## DATA ANALYSIS

This section estimates the answer reviewing effect by analyzing the common education data set from Cizek and Wollack (2017), in particular, 4th graders' math assessment data, consisting of 53 multiple-choice items. The total sample size is 71,902, and the data set records students' first and final answers for each of the items.

First, in order to estimate the sign of the ATT of answer reviewing, two proportions, $P^j(WR)$ and $P^j(RW)$ for $j$-th item, are computed. As was discussed in Equation (8), if $P^j(WR) > P^j(RW)$, then the item-specific $ATT^j$ is positive; otherwise, it is negative. Figure 1 shows the two proportions for each item. The solid line represents $P^j(WR)$ and the dashed line represents $P^j(RW)$. In 52 of the 53 items, it was found that $P^j(WR) > P^j(RW)$, therefore, the ATT of answer reviewing is generally positive, $ATT^j > 0$. Only one item, #38, shows $P^{j=38}(WR) < P^{j=38}(RW)$, suggesting a negative ATT for that item, $ATT^{j=38} < 0$. More specifically, for this item, 595 examinees first chose wrong answers but then changed to the right answer (WR), while 1,238 examinees first chose the right answer but then changed to wrong answers (RW). This exceptional case shows that it is possible that the ATT can be negative.[9]

---

[9] The specific question content of item #38 is not available and thus why this exceptional proportion difference $P(WR) < P(RW)$ occurred is uncertain. However, although it is rare, a similar case has been occasionally found in the literature. For example, see Jeon et al.'s (2017) item #43 in their Figure 2 (p. 10).

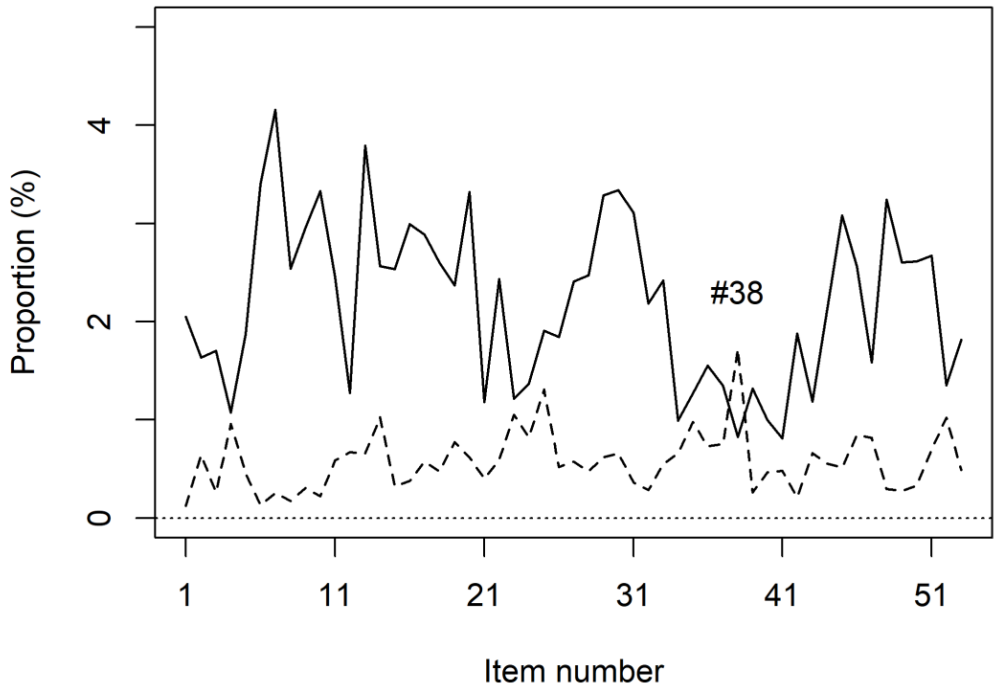

FIGURE 1. *The group proportions of WR (solid line) and RW (dashed line) for each item. The proportion of RW is greater than the proportion of WR for only one item, #38.*

Second, to estimate the sign of the ATE, the assumption-free bound in Equation (9) is applied. The results are summarized in Figure 2. All bounds, indicated by the vertical lines, are rather wide and always include zero. Thus, it is impossible to infer the sign of the ATE for any items using the answer sheet data alone. However, if one is willing to make the assumption $P^j(WW_2) > P^j(RR_1)$ for all items, then the tightened bound on the ATE in Equation (10) can be applied. Note that the lower bound is $P^j(WR) - P^j(RW)$, and this bound is indeed easily obtained in Figure 1. Therefore, under the assumption $P^j(WW_2) > P^j(RR_1)$, the ATE of answer reviewing is positive, $ATE^j > 0$ except for item #38. Note that it cannot be inferred that $ATE^{j=38} < 0$ because the bound in Equation (10) only tells us that the lower bound for #38 is negative and thus the bound will include zero. The true $ATE^{j=38}$ can be positive or negative. Overall, the real data analysis shows that the average causal effect of answer reviewing on the entire group of examinees (i.e., ATE) is likely positive.

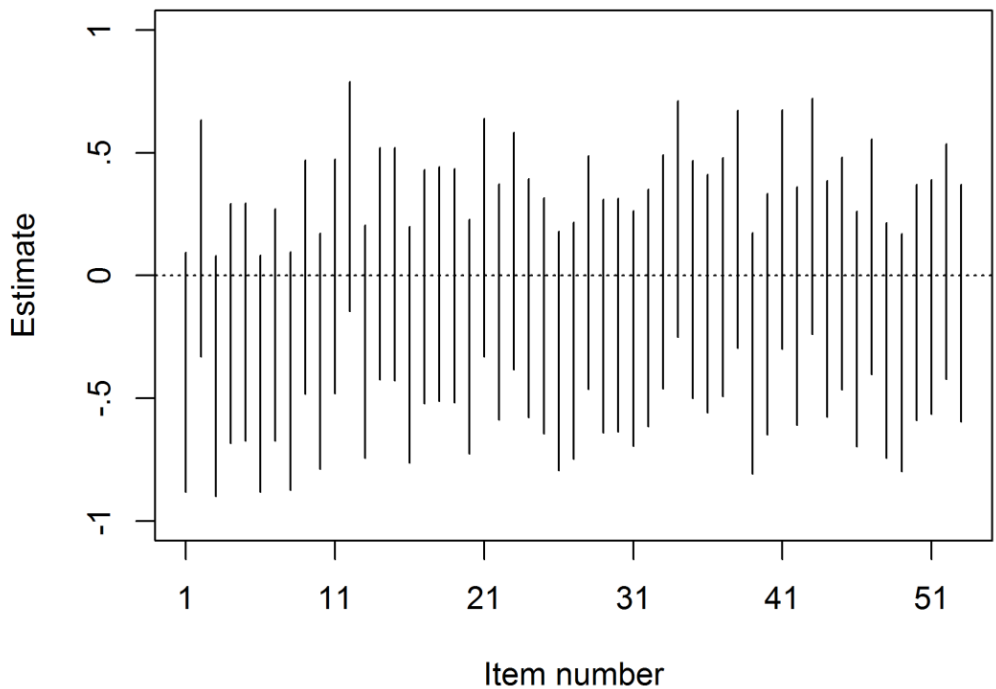

FIGURE 2. *Assumption-free bounds on ATE of answer reviewing for each item.*

As discussed, without making an assumption, the sign of $ATU^j$ cannot be empirically estimated for all items because the assumption-free bound on ATU is not informative; see Equation (11). However, if one is willing to assume that $P^j(WW_2) > P^j(RR_1)$, the ATU of answer reviewing is positive, $ATU^j > 0$, by Equation (7).

## DISCUSSION

Multiple-choice exams are the most long-standing and popular type of exams in academic and industrial settings. How examinees behave during the exams, for example, how they review and change their answers, has long intrigued measurement researchers (Benjamin et al., 1984; Lehman, 1928; Liu et al., 2015; Jeon et al., 2017; van der Linden & Jeon, 2012). This article discusses whether reviewing previous answers is beneficial for examinees to increase their final score in regular exam-taking settings. Unlike conventional measurement studies, this article investigates the question using the potential outcomes approach developed in the causal inference literature (Holland, 1986; Rubin, 1974). One of the contributions of this article is that it makes explicit what the target causal quantities are, and the conditions under which such quantities can or cannot be inferred from observed data. It separately defines the average answer reviewing effect on all of the examinees (ATE), the average answer reviewing effect on those who reviewed (ATT), and the average answer reviewing on those who did not review (ATU). Although the causal effects cannot be identified, this article develops partial identification strategies and investigates the signs of those causal effects. The main findings of this article are that, given only the answer sheet data, i) the sign of the ATT is always inferable from the data, ii) the sign of the ATE may or may not be inferred from the data, and iii) the sign of the ATU cannot be inferred from the data alone.

This formal framework helps researchers correctly interpret results in answer reviewing and changing studies. One single consistent finding in the literature is that the proportion of *WR* is generally greater than the proportion of *RW* (Benjamin et al., 1984; Liu et al., 2015; van der linden et al., 2011); this article also found that $P(WR) > P(RW)$, except in one item (see Figure 1). This finding has been mistakenly interpreted as evidence of the general benefit of answer reviewing/changing in the literature (e.g., Benjamin et al., 1984; Liu et al., 2015; also see Kim, 2019). According to the discussed formulation, this finding only indicates that the ATT of answer reviewing is positive. That is, reviewing previous answers is beneficial only to those who reviewed their answers. When combined with the additional assumption, $P(WW_2) > P(RR_1)$, which is in principle untestable, the empirical finding, $P(WR) > P(RW)$, can be interpreted as the positive ATE.[10] Remarkably, the present article infers the signs of the answer reviewing effects even though the treatment status (whether examinees reviewed their answers or not) is not fully known to researchers.

The finding of the overall positive effect of answer reviewing in regular multiple-choice exams can advance the current debate about the answer reviewing effect. Measurement researchers and test administrators have debated whether examinees should be allowed to review their previous answers in exams. Some have argued that they should because answer changing is generally beneficial (Liu et al., 2015), while others have argued that reviewing may not be necessary because answer changing can be harmful (van der Linden et al., 2011; but note that their empirical analysis remains inconclusive). Although Kim (2019) clarified that the answer reviewing effect cannot be predicted from the answer changing effect, it has remained unclear how to separately investigate the answer reviewing effect. Some prior studies performed a randomized experiment where researchers randomize examinees' reviewing status (e.g., Vispoel, 2000), but they may have an external validity issue (Shadish, Cook, & Campbell, 2002) because such lab settings are obviously different from a real exam-taking setting (e.g., the fact that a researcher forces examinees not to review their answers may change examinees' attitudes and behaviors). In contrast, this article investigates answer reviewing effects in regular exam-taking settings where examinees naturally choose whether they will review some previously answered items, reflecting their reality in the workplace.

---

[10] Making some modeling assumptions may even allow one to infer the exact magnitude of the ATE. For example, van der Linden et al.'s (2011) and Jeon et al.'s (2017) approaches allow for computing the point estimates (instead of bounds or signs) of the ATE of answer reviewing. However, in general, strong inferences (e.g., knowing the exact magnitude) require strong assumptions, which may be untestable (Manski, 2013). The proposed bounded effects do not rely heavily on strong modeling assumptions (e.g., parametric IRT models) and thus the inferred sign of the ATE of answer reviewing from the bounds is robust to the violation of such modeling assumptions.

It should be emphasized that the results of this article do not directly apply to the computer-based test condition. External validity would again be an issue because paper-and-pencil tests and computer-based tests are two very different test situations. It is not clear how each differently affects examinees' attitudes and behaviors regarding their answer reviewing decision. Nonetheless, the formal causal approach articulated in this article can be still useful because it helps researchers deal with inherently unknown counterfactuals (i.e., what would have happened if he or she had reviewed the item?), which is also the major challenge in studying answer reviewing effects in computer-based tests. This article shows that due to the special nature of multiple-choice exams, some counterfactuals can be reasonably inferred from the observed answer sheet data. In fact, partial identification of answer reviewing effects would be much easier in computer-based tests because the uncertainty caused by guessing examinees' reviewing status can be substantially reduced by the log files that record their activities during the test. Applying the approach described in this article to data sets obtained under some modified CATs that (partially) allow examinees to review their answers (e.g., the multistage adaptive test; Luecht & Nungester, 1998; the item pocket method; Han, 2013) would be a particularly interesting future topic, which shows the potential for this approach to advance the field.